\documentclass[vecphys]{svmult}

\usepackage{makeidx}         
\usepackage{graphicx}        
\usepackage{epstopdf}
\usepackage{multicol}        
\usepackage[bottom]{footmisc}

\makeindex             

\begin{document}

\title{High Time Resolution Astrophysics and Pulsars}
\author{Andy Shearer\inst{1}}
\institute{National University of Ireland, Galway
\texttt{andy.shearer@nuigalway.ie}}
%
%
\maketitle

\begin{abstract}

The discovery of pulsars in 1968 heralded an era where the temporal characteristics of detectors had to be reassessed. Up to this point detector integration times would normally be measured in minutes rather seconds and definitely not on sub-second time scales. At the start of the 21st century pulsar observations are still pushing the limits of  detector/telescope capabilities. Flux variations on times scales less than 1 nsec have been observed during giant radio pulses. Pulsar studies over the next 10--20 years will require instruments with time resolutions down to microseconds and below, high-quantum quantum efficiency, reasonable energy resolution and  sensitive to circular and linear polarisation of stochastic signals. This chapter is review  of  temporally resolved optical observations of pulsars. It concludes with estimates of the observability of pulsars with both existing telescopes and into the ELT era.


\end{abstract}

\section{Introduction}
\label{sec:2}
Traditionally astrophysics has concerned itself with minimum
time-scales measured in hours rather than seconds. This was
understandable as the available recording media were slow --- e.g. chart
recorders and photographic plates. In the 1950s, 60s and 70s wavebands
away from the optical were developed --- from ground based
radio studies to space and balloon borne high energy work. In contrast to optical wavelengths 
instrumentation in these (high and low energy) regimes was capable of
time resolutions of less than a second. Vacuum tube and CCD
technologies in the 1970s and 80s extended electronic detectors to the
optical band pass thereby allowing for studies, in the optical regime,
at all time scales. More recently superconducting detectors and
avalanche photodiodes (APD) have given us the possibility of nanosecond
observations with high quantum efficiency. To date high time
resolution observations, in optical astronomy, have been driven predominantly by detector developments and possibilities --- and not by the underlying science. This situation is now changing where studies of stochastic phenomena such as giant radio pulses (GRP) and rotating radio transients (RRAT) require specific types of detector and instrumentation.
Table \ref{astro-tim} shows the characteristic time-scale for different astronomical objects. Clearly studies of pulsars require the ability to observe down to sub-millisecond and in some cases sub-microsecond times scales. Consequently, optical observations of pulsars probably represent the biggest instrumental challenge to high-time resolution astrophysics. Radio observations \cite{hank03} have shown flux variations from the Crab during GRPs on time scales of nanoseconds. Although over 1600 radio pulsars have been observed, only five normal pulsars and one anomalous X-ray pulsar (AXP)   have been observed to pulse at optical wavelengths. Optical observations of pulsars are also limited by their intrinsic faintness in the optical regime see Table 2. 

\begin{table}
\centering
\caption{Small Timescale Variability of Astronomical Objects. Pulsars can be shown to require observations on the shortest time-scales - variability in the radio region has been shown down to nanosecond scales and in the optical variability at the microsecond scale has been observed.}
\begin{tabular}{lccc}

\hline
    &    &  \multicolumn{2}{c}{ Time-scale}  \\
    &    &  (now)   &    (ELT era) \\
    \hline
Stellar flares    &    &  ~~~~ Seconds / ~~~~&  ~~~~~10--100 ms ~~~~\\
pulsations        &     &  Minutes     &  \\
 & & & \\
Stellar surface & White Dwarfs & 1--1000 $\mu$s & 1--1000 $\mu$s \\
oscillations &       Neutron Stars &     & 0.1 $\mu$s \\
 & & & \\
Close Binary  & Tomography & 100 ms++ & 10ms+ \\
Systems           & Eclipse in/egress & 10 ms+ & $<$1ms \\
(accretion and~~~ &  Disk flickering & 10 ms & $<$1ms \\
turbulence )     & Correlations & 50 ms & $<$1ms \\
                      &   (e.g. X \& Optical) ~~~& & \\
 & & & \\
Pulsars & Magnetospheric & 1 $\mu$s--100 ms & ns(?)\\
                &  Thermal & 10 ms & $<$ms \\
 & & & \\
AGN       &             & Minutes  & Seconds \\          
\hline

\end{tabular}
\label{astro-tim}       
\end{table}

In this chapter we present an analysis of optical observations of pulsars which have shown variability on time-scales from few nanoseconds to secular changes over years. In this regard they represent the most challenging target for optical High-Time Resolution Astrophysics. Pulsars, neutron stars with an active magnetosphere, are created through one of  two main processes --- the compact core after a type II supernova explosion or as a result of accretion induced spin up in a binary system. Pulsars have strong fields up to $10^{13}$ G and rotation periods down to less than 2 milliseconds, which in combination of can produce exceptionally strong electric fields. The plasma accelerated by these fields radiates at frequencies ranging from the radio to TeV $\gamma$-rays. There is however a fundamental difference between the radio and higher energy emission --- the former is probably a coherent process and the latter some form of synchrotron or curvature radiation. Despite nearly forty years of theory and observation a number of fundamental parameters are not known :- 
\begin{itemize}
\item What is the expected distribution of the plasma within the pulsar's magnetosphere?
\item Where is the location and mechanism for accelerating the plasma?
\item Where in the magnetosphere is radiation emitted? 
\item What is the emission mechanism in the radio and at higher photon energies?
\end{itemize}

Optical observations of pulsars are limited by their intrinsic faintness in the optical regime --- early estimates indicated that the optical luminosity should scale as Period$^{-10}$ \cite{pac71} making many pulsars unobservable with current technology. However it is in the optical regime that we have two distinct benefits with regard to other high-energy wavelengths --- in the optical regime we can readily measure all Stokes' parameters including polarisation and in the optical we are probably seeing a flux which scales linearly with local power density in the observer's line of sight. With the advent of increasingly large telescopes and more sensitive detectors we should soon be in a position where the number of observed optical pulsars will have increased from five to over hundred.

\section{Normal Pulsars}
\label{sec:3}
\subsection{Observations --- The Phenomenology}

Understanding the properties and behaviour of neutron stars has been one of the longest unsolved stories in modern astrophysics. They were first proposed as an end point in stellar evolution by Baade \& Zwicky \cite{bz34} in 1934. Possible emission mechanisms from rapidily rotating magnetised neutron stars were published {\it before} \cite{pac67} their unexpected detection by Bell and Hewish in 1968 \cite{bell68}.

\begin{figure}[htbp] 
   \centering
   \includegraphics[width=5.5in]{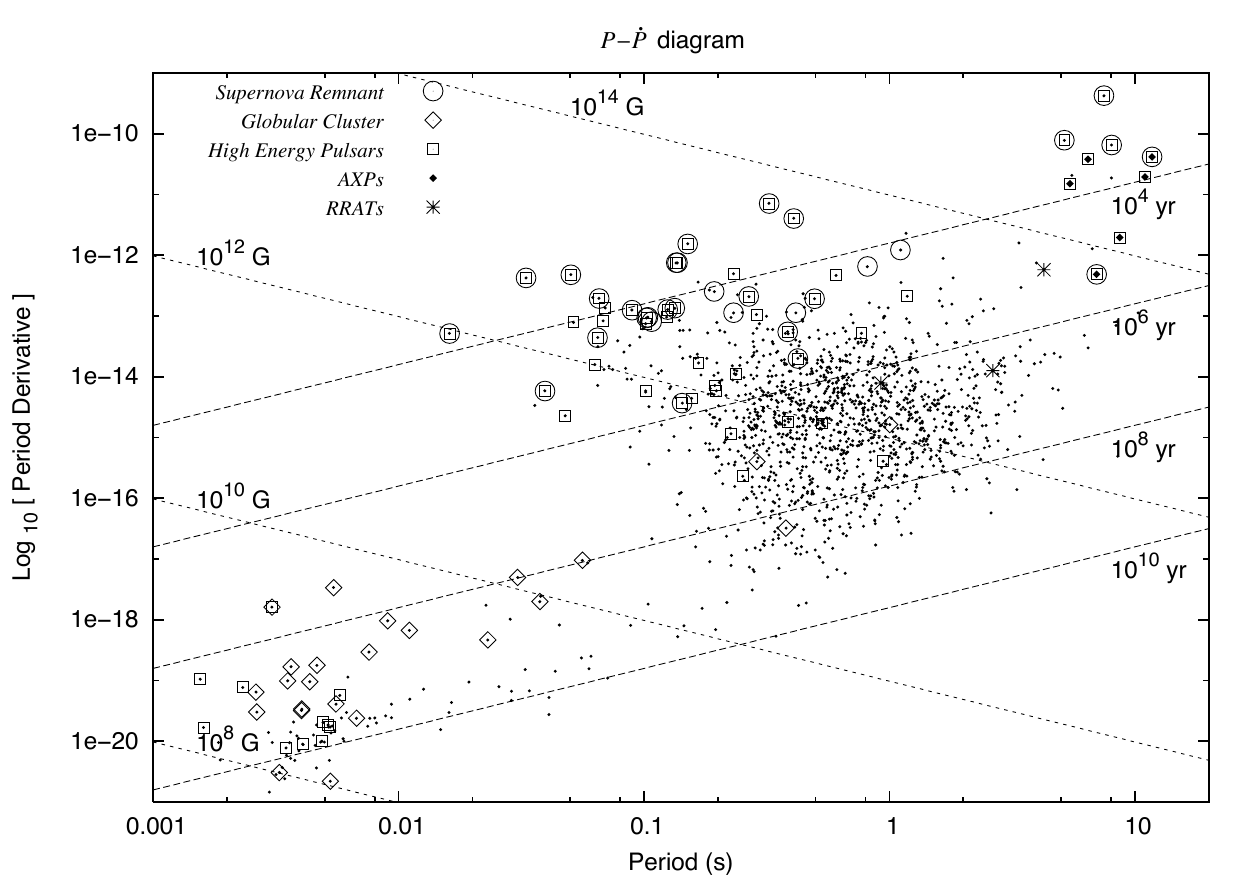} 
   \caption{The standard  $P - \dot{P}$ diagram showing the location of optical pulsars, anomalous  X-ray pulsars, rotating radio transients and pulsars associated with supernova remnants. Also marked are pulsars associated with globular clusters. The classifications are based upon data from  ATNF Catalog \cite{man05}. On the diagram we have also marked the predicted surface magnetic field and age of each pulsar. These have been determined from the normal assumption that the pulsar's spin down energy is due to magnetic dipole braking. It can be shown the the surface magnetic field, $B_S \propto (P\dot{P})^{1/2}$ and a pulsar's characteristic age as $\tau \equiv P \dot{P}^{-1}$. It should be noted that the characteristic age is normally greater than the actual age determined from historical supernovae. In this diagram normal pulsar are expected to be born towards the top left of the diagram and will move towards the bottom right. At some point they cross the {\it death line} where there is insufficient energy for $e^+e^-  $ pair creation within the magnetosphere --- this begins to explain  the lack of pulsars with long periods towards the lower right of the diagram. }
      
  \label{p-pdot}
\end{figure}

Figure \ref{p-pdot} shows the $P - \dot{P}$ diagram for pulsars which gives a gross view of  a pulsar's evolutionary position --- normal pulsars are born towards the top  of the diagram and move roughly towards the bottom right --- i.e., they slow down and it is this spin-down energy which powers the pulsar.  The optically and higher photon energy emitting pulsars can be seen to be younger and tend to have higher magnetic fields and higher $\dot{E}$ values. Millisecond pulsars, which have been spun up through accretion are found in the bottom left of diagram illustrating their relatively weak magnetic fields. The binary nature of their progenitors also increases the chance that they are found in globular clusters. 

Recently two new members of the neutron star zoo have been identified --- AXPs, now thought to be magnetars and RRATs. The former are characterised by extremely high magnetic fields ($>10^{14}$ G) and are probably powered by the decay of these fields. RRATs are the most recent neutron star observation and are characterised by their transient radio signal, which occurs for a few milliseconds, at random intervals ranging from a few to several hundred minutes. These different classes of pulsar are described below.

Since the first optical observations of the Crab pulsar in the late 1960s \cite{coc69} only four more pulsars have been seen to pulsate
optically (Vela \cite{wall77}; PSR B0540$-$69 \cite{mid85};  PSR B0656+14 \cite{shear97};  PSR B0633+17 \cite{shear98}).   Four of these five pulsars are probably too distant to have any detectable optical thermal emission using currently available technologies. For the nearest and faintest pulsar, PSR B0633+17, spectroscopic studies have shown the emission to be predominantly non-thermal \cite{mar98} with a flat featureless spectrum. For these objects we are seeing non-thermal emission, presumably from interactions between the pulsar's magnetic field and a stream of charged particles from the neutron star's surface. Four other pulsars have been observed to emit optical radiation, but so far without any detected pulsations \cite{mig05}. Optical pulsars seem to be less efficient than their higher energy counterparts with an average efficiency\footnote{The optical efficiency is defined as the optical luminosity divided by the spindown energy --- ($\eta \equiv L_\nu/\dot{E}$)} of about $10^{-9}$ compared to $\approx 10^{-2}$ at $\gamma$-ray energies \cite{lk05}. Optical efficiency also decreases with age in contrast to $\gamma$-ray emission \cite{sh01}.  These factors combined indicate that pulsars are exceptionally dim optically, although it is in the optical where it is potentially easier to extract  important  parameters from the radiation --- namely spectral distribution, flux and polarisation. 

Detailed time-resolved spectral observations have only been made of the Crab pulsar, but through broad-band photometry the spectral index of the other pulsars has been determined. Table \ref{opt-data} shows the spectral index for pulsars with observed optical emission. If the emission mechanism is synchrotron and from a single location, you would expect the spectral index to be positive for high frequencies changing to a value of 5/2  below the critical frequency in the optically thick region where synchrotron self-absorption becomes dominant. One pulsar, PSR  J0537$-$6910, which has not been observed optically despite showing many of the characteristics of the Crab pulsar, might be expected to be more luminous. It is likely that synchrotron-self absorption will play a significant part in reducing its optical flux \cite{pac71}, \cite{ocon05}. 

\begin{figure}[p] 
  
   \includegraphics[width=5in]{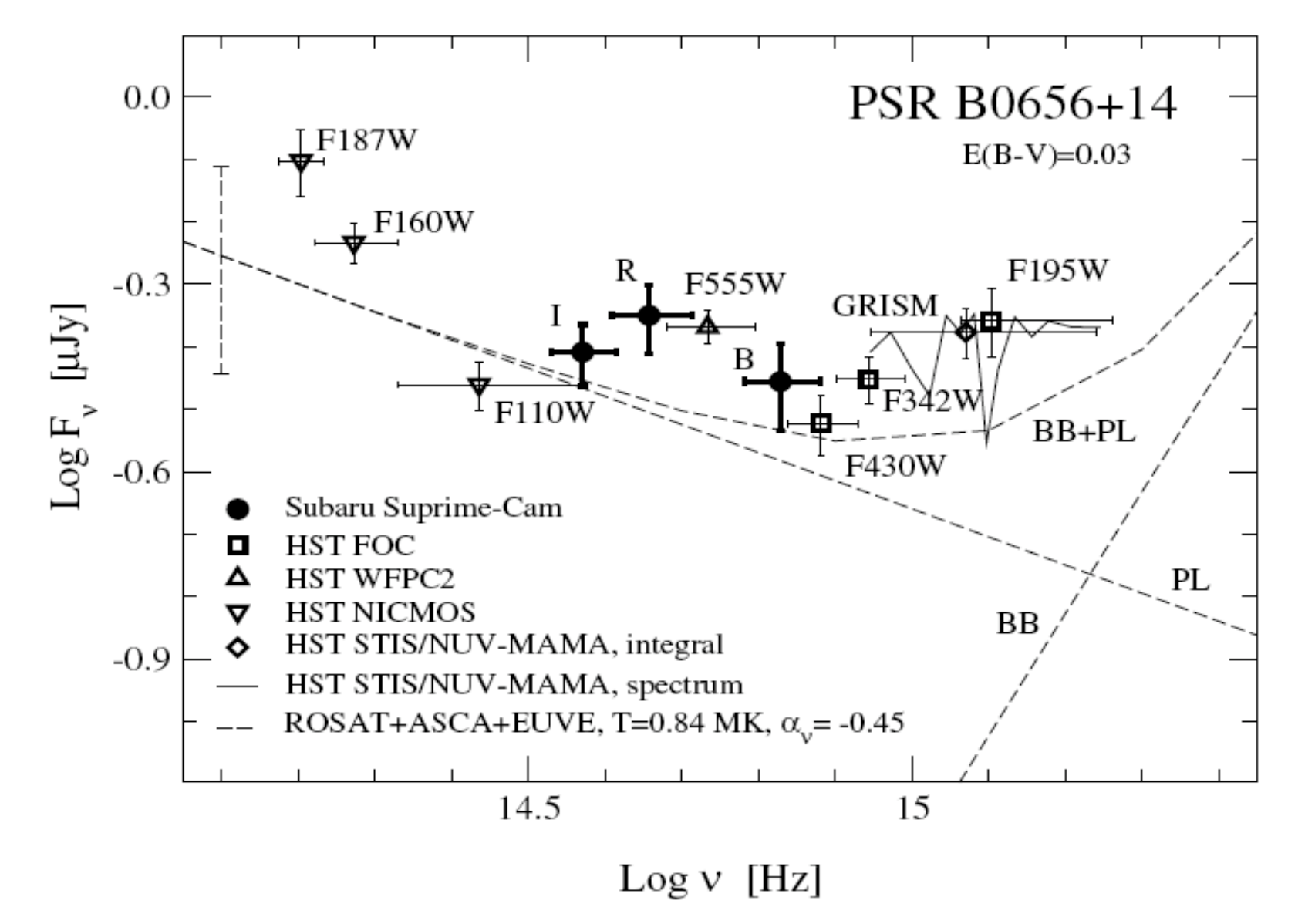} 
   \includegraphics[width=5in]{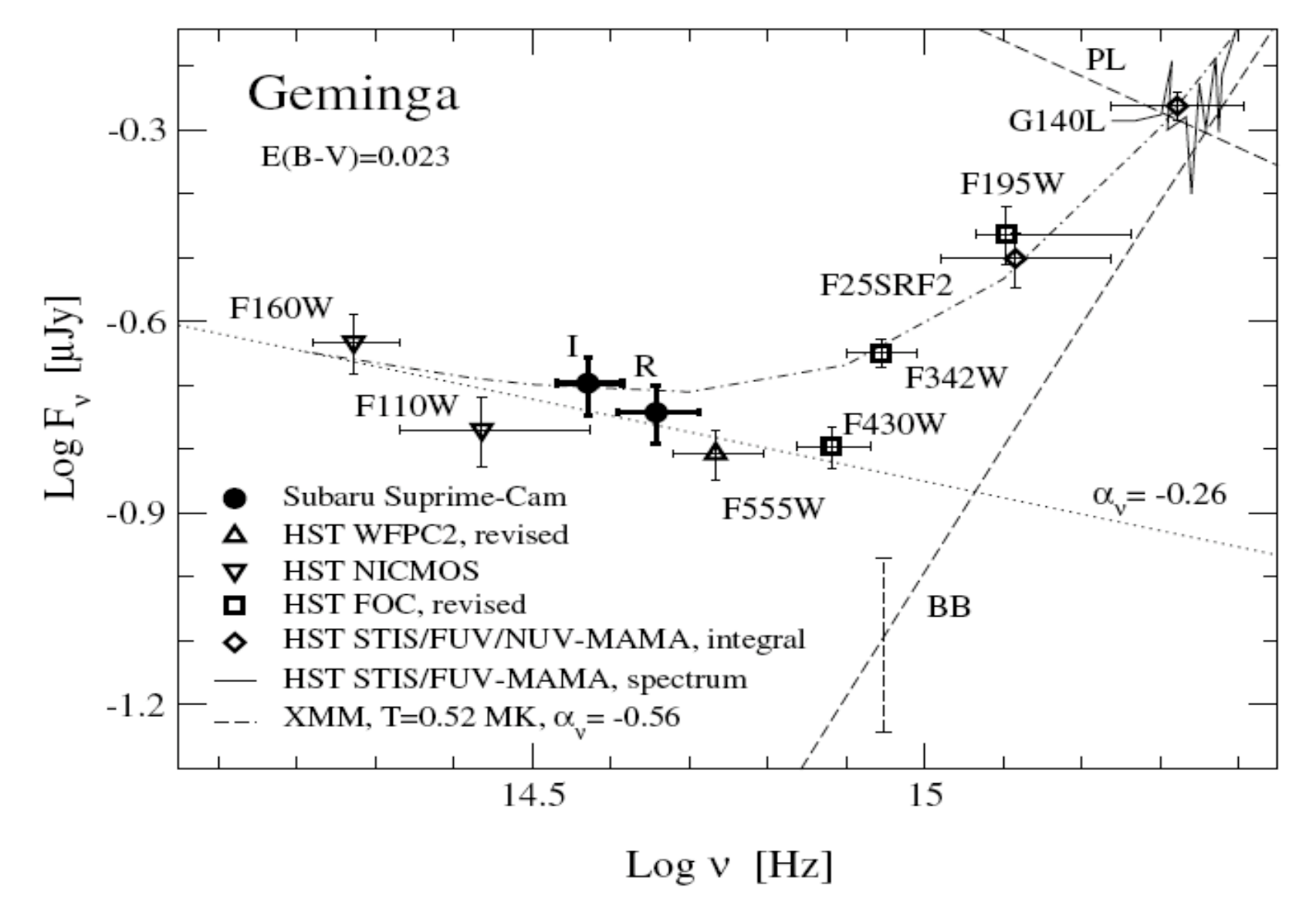} 
   \caption{Broadband photometry of the middle-aged pulsars PSR B0656+14 and Geminga showing some  similarities in spectral shape. In particular both spectra show a increase in flux towards infra-red and a possible emission feature around the V band \cite{shib06}. At the resolution of these measurements, it is not possible to determine whether this is a single feature or a number of emission lines from a pulsar wind nebula. Furthermore as these observations were time integrated it was not possible to determine how the spectral shape varied with pulse phase. }
      
  \label{sab}
\end{figure}

The other, optically fainter, pulsars have either had spectra taken with low signal-noise \cite{mar98} or using broad band photometry had their gross spectral features determined. There have been suggestions of broadband features in the spectrum possibly associated with cyclotron absorption or emission features; Figure \ref{sab} shows the IR-NUV spectrum for PSR B0656+14 and Geminga \cite{shib06}. Until spectra with reasonable resolution have been taken  we can not, with any degree of certainty, characterise the spectra in term of absorption or emission features. If a cyclotron origin for these features can be determined then we have the possibility of determining the local magnetic field strength independently of the characteristic surface magnetic field ($\propto (P\dot{P})^{-0.5}$). The overall spectral shape is consistent with the UV dominated by thermal radiation from the surface and the optical-infrared region dominated by magnetospheric non-thermal emission. Neither spectra show any evidence of synchrotron self absorption towards longer wavelengths.

\begin{table}
\centering
\caption{Optical Pulsars --- Basic Data. Pulsars names marked in bold have been observed to pulsate. the spectral index covers the region from 3500--7000 \AA ~and is of the form $F_\nu \propto \nu^{\alpha}$. The spectral index are for  time-averaged fluxes. The M31 pulsar flux  is based upon extrapolating the Crab pulsar to the distance of M31 and increasing the noise background proportionately.
}
\begin{tabular}{lrrrrr}

\hline

Pulsar & m$_B$ ~~& Period &  ~~Spectral & \multicolumn{2}{c}{ VLT   ~~~~~~  ELT}  \\
              &              &  (ms)    & Index~~ &  \multicolumn {2} {c}{~~B photons/rotation~~}  \\

\hline

{\bf Crab}               & 16.8 & 33 & -0.11 &~~~~ 3,300 & 120,000 \\
{\bf PSR B0540$-$69}   & 23    &  50& 1.6 &  17 & 610 \\
{\bf Vela }               & 24 & 89 & 0.12 &12 & 440 \\
{\bf PSR B0656+14}  & 25.5 & 385 & 0.45  & 13 & 470 \\
{\bf Geminga}  & 26 & 237 & 1.9  & 25.5 & 200 \\
M31 (Crab)  & 30-31 & 33 & - & 0.02 & 1 \\     
PSR B0950+08  & 27.1(V) & 253 & - & - & - \\
PSR B1929+10 & 25.6(V)  & 227 & - & - & - \\
PSR B1055$-$52 &  24.9(V) & 197 & $-$ &- & - \\
PSR B1509$-$58 & ~~25.7(V) & 151 & - & - & - \\ 
\hline
\end{tabular}
\label{opt-data}       
\end{table}

Only one pulsar, the Crab, has had detailed polarisation measurements made \cite{sm88}, \cite{rm01a}. The polarisation profile shows emission aligned with the two main peaks and is consistent with synchrotron emission. One unusual feature is the location of maximum polarisation which for the Crab pulsar precedes the main pulse at a phase when the optical emission is at a minimum. The polarisation of one other, PSR B0656+14 \cite {kern03} has been observed albeit at low significance. This pulsar has maximum polarisation coincident with the maximum luminosity. Three other pulsars have only had their time average polarisation measured in the optical regime \cite{wag00}.

\subsection{Emission Theory}

Many suggestions have been made concerning the optical emission process for these young and middle-aged pulsars. Despite many years of detailed theoretical studies and more recently limited numerical simulations, no convincing models have been derived which explain all of the high energy properties. There are similar problems in the radio, but as  the emission mechanism is radically different (being coherent) only the high energy emission is considered here. In essence despite nearly forty years of observations, we still do not understand the emission mechanisms behind pulsar emission. The various competing theories have all failed to provide a comprehensive description of pulsar emission. Ò{\it Probably the only point of agreement between all these theories is the association of pulsars with magnetized, rotating neutron stars}Ó \cite{ly99}.  In the high energy regime there is more consensus as to the emission mechanism --- either incoherent synchrotron or curvature radiation for outer gap models or inverse Compton scattering for polar gap models. However, like radio emission there is not a consensus as to the location of the emission region or the acceleration mechanisms for the plasma.

\subsection{Crab Pulsar}

The Crab pulsar has been observed for nearly forty years. The earliest photometric observations \cite{ne69}, \cite{kr70} gave a visual magnitude of $m_V ~ 16.5 \pm 0.1$ compared to more recent observations \cite{nas96} of $m_V ~ 16.74 \pm 0.05$ in the same period the pulsar has slowed by $\approx$1.5\%. If we assume a $L\propto P^{-10}$ scalling law then in the same period the luminosity should have reduced by about 0.15 magnitudes --- in reasonable agreement with the observed luminosity decrease.

As can be seen from Table \ref{opt-data} the Crab Pulsar is uniquely bright amongst the small population of optical pulsars. It is reasonably close ($\approx$ 2 kpc) and less than one thousand years old. It is the only pulsar which is bright enough for individual pulses to be observed with any significant flux - allowing for pulse-pulse spectral and polarimetric changes to be observed.  As a pulsar it has unusual characteristics --- for example it glitches and the radio emission is dominated by giant radio pulses (GRPs). The latter occur sporadically with a mean repetition rate of about 1 Hz --- the highest of the GRP emitting pulsars --- see below. The Crab pulsar is seen to emit at all energies --- from radio to high-energy $\gamma$-rays with E $>$ 1 TeV. In the optical (UBV) regime the emission is characterised by a flat power law $F_\nu \propto \nu^{-0.07 \pm 0.18}$\cite{gol00}. Although no direct evidence of a roll-over in the spectrum has been observed there are hints that in the near infra-red that the slope steepens (\cite{ocon05} and references therein). Recent Spitzer observations \cite{te06} possibly show the beginning of a rollover at $\approx$2 $ \mu$m. More detailed observations, particularly time-resolved from 1--50 $\mu$m will be required to confirm this.   
    
  \begin{figure}[htbp] 
   \centering
   \includegraphics[width=5in]{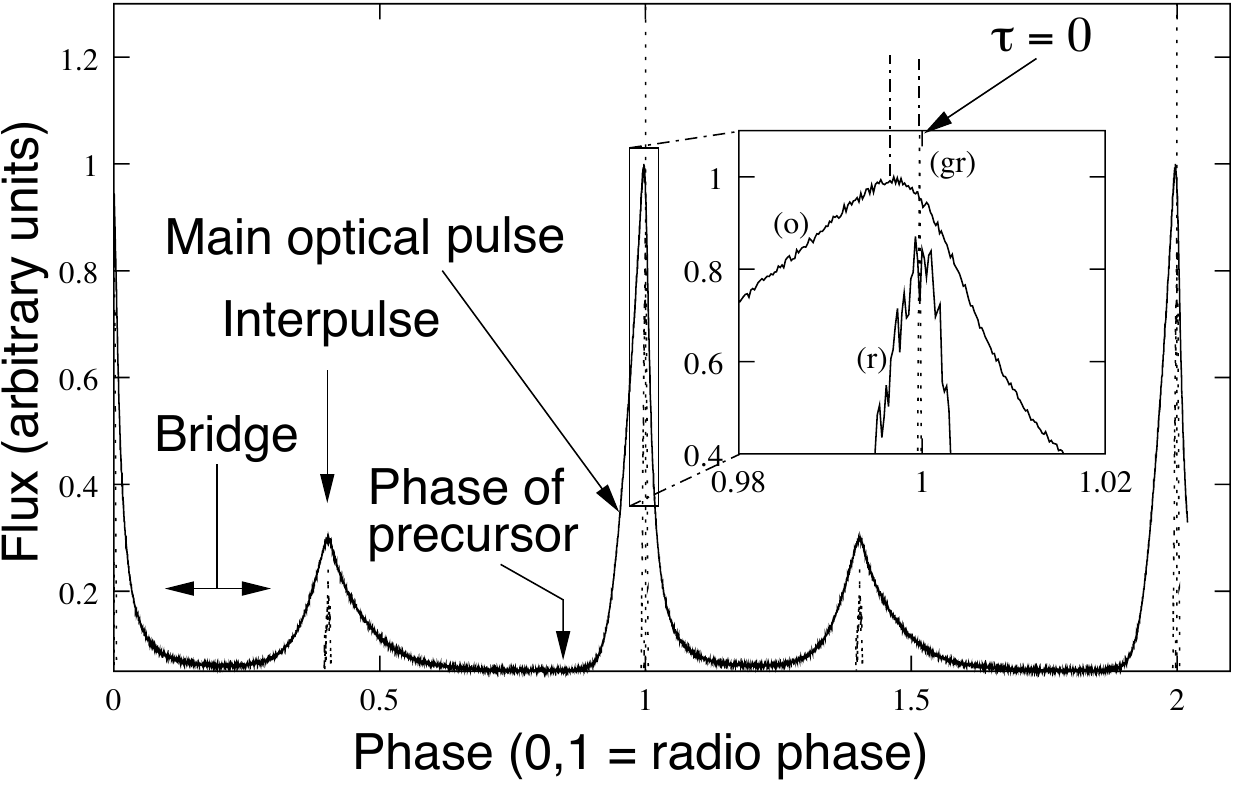} 
   \caption{Crab pulsar's light curve showing the main emission regions and the relationship between the  optical light curve and its radio counterpart \cite{shear03}. The optical data was taken using an APD based photometer on the William Herschel Telescope (WHT) each photon's arrival time has been folded in phase with the pulsar radio ephemeris to produce the time averaged light curve. The radio data has been treated in a similar fashion. Also shown is the location of a single giant radio pulse.}
   \label{crab}
\end{figure}

The Crab pulsar's light curve, Figure \ref{crab}, shows four distinct regions
\begin {itemize}
\item Main pulse containing about 50 \% of the optical emission, the pulse has a FWHM of 4.5 \% of the pulsar's period
\item Interpulse with about 30 \% of the emission 
\item Bridge with about 20 \% of the emission
\item Off-pulse just before the main pulse where  the polarisation percentage is at a maximum
  \end{itemize}
  
Spectral observations, of the Crab pulsar --- both pulsed and time-integrated --- have been made by a number of groups. All authors report spectra that is generally flat and featureless --- consistent with a synchrotron origin for the radiation with an integrated spectral index of $0.11$. A possible cyclotron absorption feature has been reported \cite{nas96} although not seen by other observers. Of note are time resolved observations which indicate a change in the spectral index on the leading and falling edge of the main and secondary peaks \cite{rm01a}, \cite{for02}, \cite{eik97}. As yet there is no clear explanation for either spectral shape changes, pulse width variations or polarimetry.

\subsection{Giant Radio Pulse Pulsars --- optical considerations}
\label{sec:5}

Six pulsars, see Table \ref{grp}, exhibit giant radio pulse phenomena. Here, as well as the normal radio emission, we see enhanced radio emission up to several times the average occurring for a few percent of the total number of pulses. The Crab pulsar was itself discovered through its giant pulse phenomena \cite{st68}. Phenomenologically these pulsars are a very diverse group with the only possible common factor being their magnetic field strength at the light cylinder  --- the pulsars in Table \ref{grp} are all in  the top six ranked by magnetic field strength at the light cylinder. The sixth pulsar in the list of pulsars with high magnetic field strength at the light cylinder, PSR J0537$-$6919 has only been observed at X-ray energies and hence no giant radio emission. Two of the pulsars, the Crab  and PSR J0540$-$6919, have very similar properties - young [age $<$ 1000 years]  pulsars embedded in a plerion. The other three are all millisecond pulsars. 

Observations of GRP pulsars at other energies has been difficult - primarily due to the low rate of GRP events (ranging from $10^{-6}-2$ Hz) . $\gamma$-ray observations \cite{lun95} showed an upper limit of 2.5 times the average $\gamma$-ray flux based upon 20 hours of  simultaneous radio-$\gamma$-ray observations with CRGO/OSSE  and the Greenbank 43m telescope. Optical observations \cite{shear03} indicated a small ($\approx$3\%) increase in the optical flux during the same spin cycle as a GRP. Although the percentage increase was small the energy increase in the optical pulse is comparable to the energy in the GRP.  These results linked for the first time flux changes in the radio --- where the emission is highly variable --- to higher energy emission which is seen to be stable. Furthermore, given the different emission mechanisms [coherent versus incoherent]  such a correlation was unexpected. Although small, the energy in the optical and radio GRPs are similar; if this were extrapolated to $\gamma$-ray energies we would expect an enhancement of $<$0.01\%.

 In two of these, PSR J1824$-$2452 and PSR J1939+2134, the GRP emission aligns in phase with the X-ray pulse rather than the normal radio pulse \cite{rm01b}\cite{cus03}. This possibly implies a different emission zone and/or mechanism. Furthermore the  spectrum of GRP fluxes shows a different slope to the normal emission and cannot be regarded simply as its high flux extension. What characterises a GRP event compared to the high-energy tail of the normal distribution is very subjective. An acceptable definition of what is and what is not a GRP is that the pulses show a power-law energy distribution (cf. lognormal for typical pulsar emission) and have very short time-scales (typically $\approx$ of nanosecond duration) \cite{rm01b}. Most studies have used a 10 or 20 $\langle E \rangle$  criteria for GRPs.

From an optical observational perspective GRPs are significantly more difficult than normal pulsars to observe --- the pulses arrive randomly albeit in phase with normal emission and at a rate significantly less than the normal pulsar rate. The random nature of GRP arrival times make synchronised systems, such as clocked CCDs, inappropriate. Polarisation studies of these type of event also restrict the type of polarimeter which can be used in the GRP studies, and makes the simultaneous measurements of all Stokes' parameters essential.

\begin{table}
\centering
\caption{Giant radio pulsars}
%
%
\begin{tabular}{lrrrcl}
\hline
Name & Period & Surface Field & Light Cylinder  & GRP & Ref. \\
              &      (ms)      &    ($10^{9}$ G)   & Field ($10^{6}$G) & Rate (Hz) & \\
\hline
Crab                        &   33.1  &          3800    &  9.8  & 1-2 & \\
PSR J0540$-$6919  &   50.4  &          5000    &  3.7 & 0.001 & \cite{jo03}\\
PSR J1824$-$2452  &      3.1  &         2.2      &  0.7 & 0.0003 & \cite{jo01} \\
PSR J1959+2048  &      1.6  &          0.2     &  1.1 & & \\
PSR J1939+2134  &      1.6  &         0.4      &  1.0 & 0.0001& \cite{so04} \\
PSR J0218+4231 &              &                    &         & 0.03 & \\

\hline
\end{tabular}
\label{grp}
\end{table}

\begin{figure}[htbp] 
   \centering
   \includegraphics[width=5in]{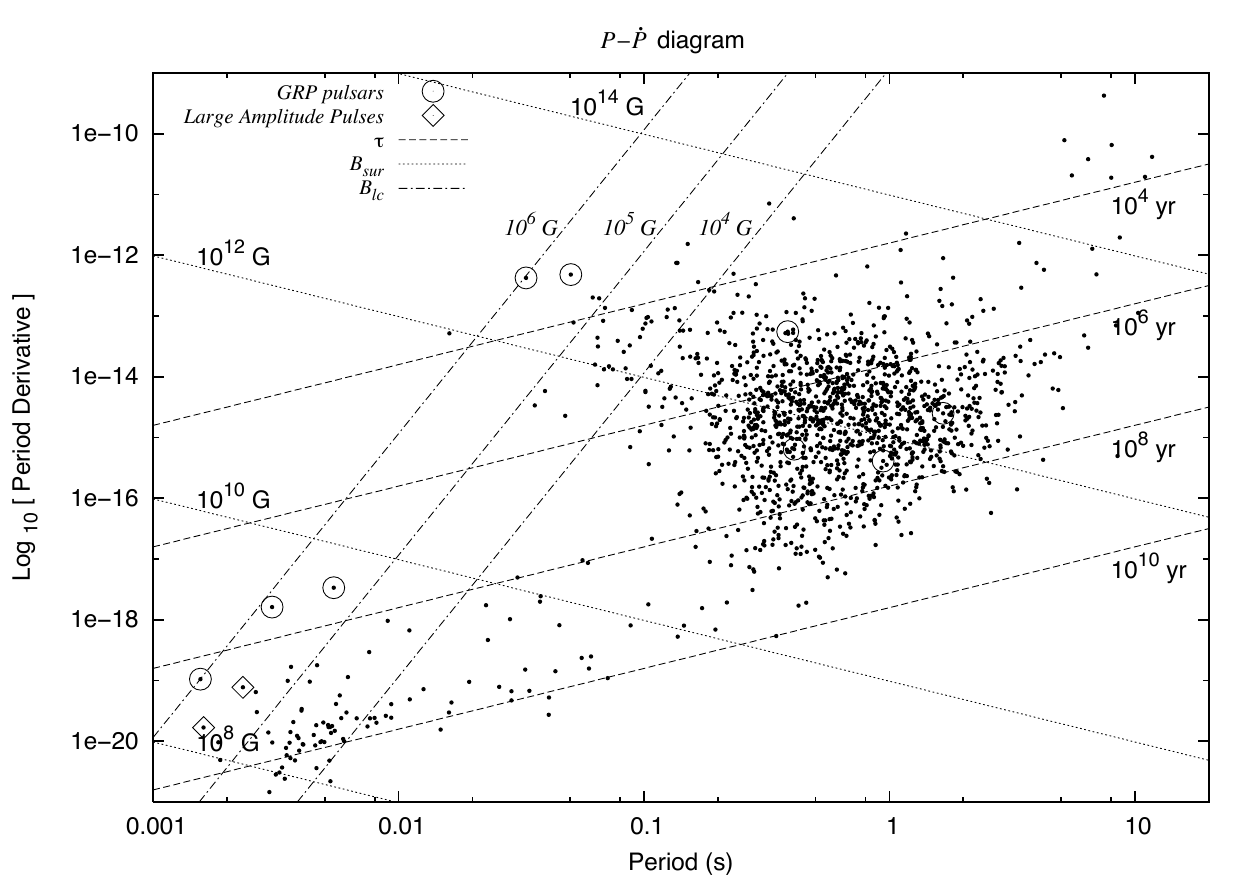} 
   \caption{The location of giant pulse emitters on the $P-\dot{P}$ diagram. We have also shown the inferred maximum magnetic field at the light cylinder based upon a 10 km radius neutron star. }
   \label{fig:grp}
\end{figure}

\section{AXPs and RRATs}

\subsection{Anomalous X-Ray Pulsars}
\label{sec:4}

Anomalous X-Ray pulsars are a special class of neutron star characterised by very high inferred magnetic fields and variable X-ray emission. Soft Gamma-ray repeaters (SGRs), 
first observed as transient $\gamma$-ray sources and now known to be persistent sources of pulsed X-rays, show similar properties. Both AXPs and SGRs are thought to have similar 
physical properties and are likely to be magnetars - where the emission comes from the decay of the strong magnetic field rather than being rotation powered. The review by Woods and
 Thompson \cite{wt04} is a recent review of general AXP and SGR properties.

In Figure \ref{p-pdot}  AXPs are shown to be towards the top right of the $P-\dot{P}$ diagram. Models for their emission mechanism historically concentrated upon the interaction between the neutron star and an accretion disk.  However the lack of a secondary star precluded all but a fossil disk around the star. The latter explanation was precluded by infrared observations \cite{hull00}. The faintness of the infrared counterpart precluded a disk model and strengthened the case for the magnetar model where the high-energy emission comes from the decay of the strong magnetic field or from other magnetospheric phenomena.

Optical pulsed AXP emission was first observed from 4U 0142+61, using a phase-clocked CCD \cite{ker02} and confirmed by UltraCam observations \cite{dhill05}. From Table \ref{table-axp} it can be seen that this pulsar is significantly brighter than other AXPs in the infra-red and all suffer from significant reddening. All of which combines to make future AXP optical observations very difficult and dependent on system such as Ultracam, but with better infra-red sensitivity.

\begin{table}
\centering
\caption{Optical and Infra-red properties AXPs and SGRs}
\begin{tabular}{llcccccc}
\hline
Source              & P   & $\dot{P}$                  & B                      &  \multicolumn{3}{c}{  Magnitude} \\ 
                           & (s) & $10^{-11} s s^{-1} $ & $10^{14}$ G & V & I & J  \\
SGR 0526$-$66 &   8.0    &     6.6                      &   7.4                & $>$ 27.1 & $>$25 & - \\
SGR 1627$-$41 &    6.4   &                                 &                         &  & & $>$21.5 \\
SGR 1806$-$20 &    7.5   &      8.3-47                &   7.8                      &  & & $>$21\\
SGR 1900+14 &    5.2   &      6.1-20              &     5.7                    &  & &  $>$ 22.8\\
4U 0142+61 & 8.69 & 0.196  & 1.3   & 25.6 & 23.8 & - \\
1E 1048Ð5937 & 6.45 & 3.9 &  3.9   &  & 26.2 & 21.7 \\
RXS 1708Ð4009  & 11.00 & 1.86 & 4.7 & - & - & 20.9 \\ 
1E 1841Ð045  &  11.77 & 4.16 & 7.1 &  $>$ 23(R) & - & - \\
1E 2259+586 & 6.98 & 0.0483 & 0.60 & $>$26.4(R) & $>$25.6 & $>$23.8 \\
AX J1845.0Ð0258 & 6.97 & - & - & - & - \\
CXOU J0110043.1 & 8.02 & - & - & - & - \\
XTE J1810Ð197  & 5.54  & 1.15& 2.9  & - & $>$24.3  & \\
\hline

\end{tabular}
\label{table-axp}       
\end{table}

\subsection{Rotating Radio Transients}

Rotating Radio Transients (RRATs) are a new class of pulsar that exhibit sporadic radio emission lasting for few milliseconds (2-30ms) at random intervals ranging from a few hundred seconds to several hours. Table \ref{table-rrat}, based upon \cite{mclaugh06} details the basic RRAT parameters. For a few RRATs a periodicity and period derivative has been established by using the largest common divisor to estimate the period and in three cases the period derivative. It is not known where these lie in the now expanding pulsar menagerie,  however one RRAT lies towards the magnetar region of the $P-\dot{P}$ diagram - see Figure \ref{p-pdot}. 

Todate no RRAT has had any optical counterpart observed,  however UltraCam observations of PSRJ1819$-$1458 \cite{dhill06} produced an upper limit of 3.3, 0.4 and 0.8 mJy at 3560, 4820 and 7610 \AA, respectively in 1800 seconds on the WHT. The characteristics of the emission is difficult to determine at this stage although the bursts do not seem to have the same form as Giant Radio Pulses, for example they do not show a power law size distribution . One suggestion is that we are seeing a selection effect (\cite{we06}) --- PSR B0656+14 would appear as an RRAT if it was located at a distance of greater than 3kpc. The spread of derived periods is from 0.4-8 seconds which when combined with the inferred surface magnetic field for one object being as high as $5~10^{13}$G we have a possible link to AXPs and magnetars.

\begin{table}
\centering
\caption{Observational Properties of Rotating Radio Transients}

%
%
\begin{tabular}{llcccc}
\hline
Name & Period & $\dot{P}$                 & Distance   &  Rate & On\\
            &    (s)     &  $10^{-15} s s^{-1}$ & (kpc) & (hr$^{-1}$) & Time ($10^{-5}$) \\ 
\hline
J0848$-$43 & 5.97748 & -  & 5.5  &  1.4 & 1.2\\
J1317$-$5759 & 2.6421979742 & 12.6 & 3.2 & 4.5 & 1.3 \\ 
J1443$-$60 & 4.758565 & - &  5.5 & 0.8 & 0.4 \\
J1754$-$30 & 0.422617 & Ð & 2.2 & 0.6 & 0.3 \\
J1819$-$1458 & 4.263159894 & 50.16 & 3.6 & 17.6 & 1.5  \\
J1826$-$14 & 0.7706187 & Ð & 3.3 & 1.1 & 0.06\\
J1839$-$01 & 0.93190 & Ð & 6.5 & 0.6 & 0.3 \\
J1846$-$02 & 4.476739 & Ð & 5.2 & 1.1 & 0.5 \\
J1848$-$12 & 6.7953 & Ð & 2.4 & 1.3 & 0.07 \\
J1911+00 & & - & 3.3 & 0.3 & 0.04 \\
J1913+1333 & 0.9233885242  & 7.87 & 5.7 & 4.7 &  0.3 \\

\hline
\end{tabular}
\label{table-rrat}
\end{table}

\section{Future Observing Campaigns}
\begin{figure}[htbp] 
   \centering
   \includegraphics[width=5in]{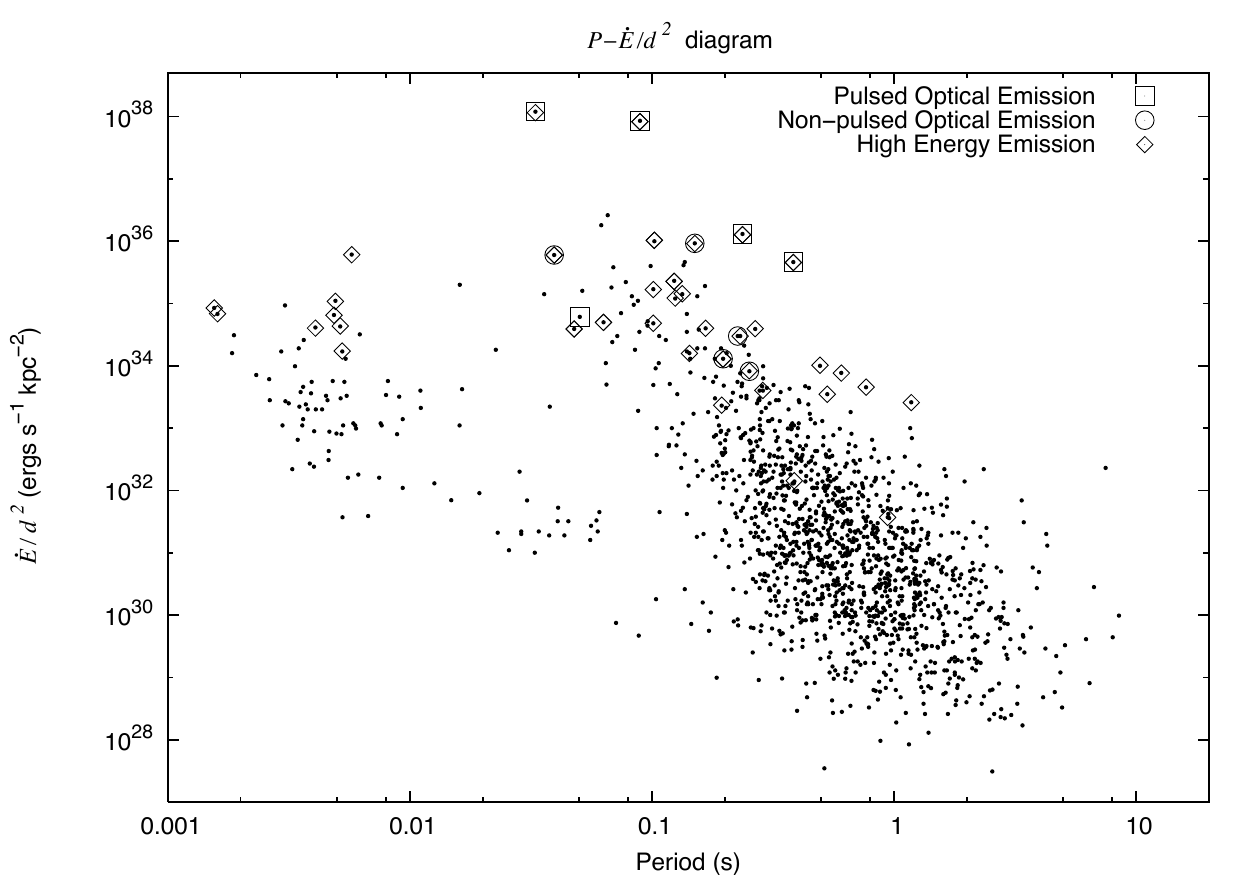} 
   \caption{Observability diagram}
   \label{obs}
\end{figure} 

Despite forty years of of observation there are still a number of fundamental unanswered question in pulsar astrophysics. In the optical regime we can be begin to answer these question if we have a larger and more comprehensive survey of normal pulsars. Most pulsed observations were made on medium sized (4-6 m) telescopes using detectors with modest or low quantum efficiencies. By moving to larger 8-10 m class telescopes we gain about a magnitude in our upper limits. By using more efficient detectors another two magnitude improvement is possible making pulsed observations of 27th magnitude objects plausible in one night. Figure 4 shows a measure of the observability (ranked according to flux)  ($ \dot{E}/d^2$) of known pulsars. If we take, an admittedly arbitrary, level of $10^{35}$ ergs sec$^{-1}$ kpc$^{-2}$ we have 20-30 pulsars which could be observed using current technology. From Figure \ref{obs} there is a rough divide around pulsar periods around 50-100 ms below which non-CCD technologies (Superconducting Tunnel Junction (STJ) devices \cite{per99}, Transition Edge Sensors (TES) \cite{rom03}, APDs) are appropriate.  In all cases polarisation measurements will be important as they give information of the local magnetic field strength and geometry. Spectral information, to look for synchrotron self absorption and cyclotron features, will give important clues to the strength of the magnetic field in the emission zone and thus its height  above the neutron star surface. In the future coordinated campaigns, particularly simultaneous radio-optical observations, can be used to look for specific phenomena such as GRPs and RRATs. 

\vspace{2mm}

{\bf Normal Pulsars}
\vspace{1mm}

Only five normal pulsars have had observed pulsed optical emission. From such a small sample it is not possible, particularly with the lack of an acceptable theory for pulsar emission, to make any strong conclusions. However with high quantum efficiency detectors and larger telescopes we can reasonably expect an increased sample of optical pulsars over the next 5-10 years. Specifically if we take the optical luminosity to scale with the spin down energy ( $L_O \propto \dot{E}^{1.6}$) \cite{sh01} then we can estimate the number of pulsars which are potentially observable. This is a very approximate relationship designed to indicate which pulsars are more likely to observed with current telescopes and instrumentation. From Table \ref{opt-pul} the following pulsars can be potentially observed with 4-8m class telescopes with current optical detectors --- PSR J0205+6449, PSR J1709$-$4429,  PSR J1513$-$5908,  PSR J1952+3252, PSR J1524$-$5625, PSR J0537$-$6910 and PSR J2229+6114.

\begin{table}

\centering
\caption{Pulsar observability table, based upon the ATFN Pulsar Catalog\cite{man05}, showing the top twenty normal, non-binary pulsars ranked according to $\dot{E}/d^2$. Only those pulsars with  $\dot{E}/d^2 > 10^{35}$ and those with observed X and $\gamma$-ray emission have been included. In this we have not made any asumptions about the efficiency of optical emission with age, field, period or any of the normal phenomenological parameters. Where not explicitly referenced the A$_V$ values are based upon $N_H=1.79 \pm 0.03 A_V 10^{21} $cm\cite{ps95}, values in italics have been estimated from this relationship. The predicted magnitude is based upon the transverse light cylinder magnetic field,  Luminosity $\propto \dot{E}^{1.6}$ \cite{sh01}. This relationship should be treated with caution and is purely phenomenological. The aim here is to give a rough scaling between the different pulsars. No millisecond pulsar have been included in this table. Table 7 shows the similar table for millisecond pulsars without any prediction of optical luminosity. }

\begin{tabular}{lrcccccccll}
\hline

Name & Period  & Distance   & log(age) & log($\dot{E}$ & log($\dot{E/d^2}$) & N$_H$ &$A_V$  &  Pred. & Association& Ref.\\
            &    (ms)  & (kpc)         & years         &  ergs/s         &     ergs/s/kpc$^2$ &      $10^{22}$cm$^{-2} $    &             & V Mag. & & \\ 
\hline
J0534+2200 & 0.0331 & 2 & 3.09 & 38.66 & 38.06 & 0.3 & 1.6 & {\bf 16.8} & M1 & \cite{bt97}\\
J0835$-$4510& 0.0893 & 0.29 & 4.05 & 36.84 & 37.93 & 0.04 & 0.2 &  {\bf 24} & Vela & \cite{bt97}\\
J0205+6449 & 0.0657 & 3.2 & 3.73 & 37.43 & 36.42 & 0.3 & 1.7 & 23 & 3C58 & \cite{hef95} \cite{tor00} \\
J1833$-$1034 & 0.0619 & 4.3 & 3.69 & 37.53 & 36.26 & 2.2 & 12.3 & 34 &G21.5-0.9 &  \cite{saf01} \cite{cam06} \\
J0633+1746 & 0.2371 & 0.16 & 5.53 & 34.51 & 36.12 & 0.01 & 0.07 &  {\bf 26} &   & \cite{bt97} \\
J1709$-$4429 & 0.1025 & 1.82 & 4.24 & 36.53 & 36.01 & 0.54 & 3.0 & 26.5 & G343.1-2.3 & \cite{bt97}\\
J1513$-$5908 & 0.1507 & 4.4 & 3.19 & 37.25 & 35.96 & 0.8 & 5.2 & 28 & G320.4-1.2 & \cite{bt97}\\
J1952+3252 & 0.0395 & 2.5 & 5.03 & 36.57 & 35.78 &  0.34 & 1.9 & 26 & CTB80 & \cite{bt97}\\
J1930+1852 & 0.1369 & 5 & 3.46 & 37.06 & 35.66 & 1.6 & 8.9 & 33 & G54.1+0.3 & \cite{lu02} \\
J0659+1414 & 0.3849 & 0.29 & 5.05 & 34.58 & 35.66 & 0.01 & 0.09 &  {\bf 25.5} & Mon. Ring & \cite{bt97}\\
J1124$-$5916 & 0.1353 & 5.4 & 3.46 & 37.08 & 35.61 & 0.31 & 1.7 & 25 & G292.0+1.8 & \cite{hug03}\\
J1747$-$2958 & 0.0988 & 2.49 & 4.41 & 36.40 & 35.61 & 2-3 & 1.4 & 26 & G359.23-0.82 & \cite{cam02} \cite{ga03} \\
J1617$-$5055 & 0.0694 & 6.46 & 3.91 & 37.20 & 35.58 & 6.8 & 38 & 62 & RCW 103  & \cite{tor98} \cite{got97} \\
J1048$-$5832 & 0.1237 & 2.98 & 4.31 & 36.30 & 35.35 & 0.5 & 2.8 & 28 & &  \cite{bt97} \cite{piv00} \\
J1524$-$5625 & 0.0782 & 3.84 & 4.50 & 36.51 & 35.34 & - & - & *  &  & \cite{kr03}\\
J0537$-$6910 & 0.0161 & 49.4 & 3.69 & 38.69 & 35.30 & 0.1-0.6 & 2 & 24 & N157B & \cite{ma98}\cite{to06} \\
J1357$-$6429 & 0.1661 & 4.03 & 3.86 & 36.49 & 35.28 & - & - & * &G309.8-2.6? & \cite{ca04} \\
J1420$-$6048 & 0.0682 & 7.69 & 4.11 & 37.02 & 35.24 & 2.2 & 12.3 & 37 &  Kookaburra & \cite{ro01} \\
J1826$-$1334 & 0.1015 & 4.12 & 4.33 & 36.45 & 35.22 & 8.2 & 46 & 71 & & \cite{bt97}\\
J2229+6114 & 0.0516 & 12.03 & 4.02 & 37.35 & 35.19 & 0.6 & 3.4 & 28 & G106.6+2.9 & \cite{ha01} \\
J1913+1011 & 0.0359 & 4.48 & 5.23 & 36.46 & 35.16 & - & - & * & & \cite{kr03} \\
J1803$-$2137 & 0.1336 & 3.94 & 4.20 & 36.35 & 35.16 & 14.0 & 78 & 104 & G8.7-0.1(?) & \cite{bt97}\\
J1809$-$1917 & 0.0827 & 3.71 & 4.71 & 36.25 & 35.11 & - & - & * & & \cite{kr03} \\
J1740+1000 & 0.1541 & 1.36 & 5.06 & 35.37 & 35.10 & - & - & * & & \cite{mc02}  \\
J1801$-$2451 & 0.1249 & 4.61 & 4.19 & 36.41 & 35.09 & - & -& * & G5.4-1.2 & \cite{kr03} \\
J0940$-$5428 & 0.0875 & 4.27 & 4.63 & 36.29 & 35.03 & - & - & *  & & \cite{kr03} \\
J0540$-$6919 & 0.0504 & 49.4 & 3.22 & 38.17 & 34.78 & 0.46 & 0.62 & {\bf 23} & SNR 0540-693 & \cite{ser04} \\
J1105$-$6107 & 0.0632 & 7.07 & 4.8 & 36.39 & 34.69 & 0.7 & 3.9 & 31& & \cite{go98} \\
J1932+1059 & 0.2265 & 0.36 & 6.49 & 33.59 & 34.48 & 0.04 & 0.2 & 32 & &  \cite{bec06} \\
J0538+2817 & 0.1432 & 1.77 & 5.79 & 34.69 & 34.20 & 0.3 & 1.7 & 33 & S147 & \cite{ro03} \\
J1057$-$5226 & 0.1971 & 1.53 & 5.73 & 34.48 & 34.11 & 0.12 & 0.7  & 32 & &  \cite{mig97} \\
J0953+0755 & 0.2531 & 0.26 & 7.24 & 32.75 & 33.91 & 0.03 & 0.2  & 35 & & \cite{za04} \\
J0631+1036 & 0.2878 & 6.55 & 4.64 & 35.24 & 33.61 &  0.2 & 1.1 & 33 & & \cite{to00}\\
J0826+2637 & 0.5307 & 0.36 & 6.69 & 32.66 & 33.55 & 0.08 & 0.5  & 36 & & \cite{be04} \\

\hline
\end{tabular}
\label{opt-pul}

\end{table}

\begin{table}
\centering
\caption{Millisecond pulsar observability table, based upon the ATFN Pulsar Catalog \cite{man05}} 


\begin{tabular}{lrccccc}
\hline
Name & Period  & Distance   & log(age) & log($\dot{E}$ & log($\dot{E/d^2}$) & Ref.\\
            &    (ms)  & (kpc)         & years         &  ergs/s         &     ergs/s/kpc$^2$ &    \\ 
\hline
J2124$-$3358  & 4.93 & 0.25 & 9.6 & 33.8 & 35.0 &    \cite{be00} \\
J1824$-$2452  & 3.05 & 4.9 & 7.5 & 36.4 & 35.0 &   \cite{bt97}\\
J1939+2134  & 1.56 & 3.6 & 8.4 & 36.0 & 34.9 &      \cite{bt97}  \\
J0030+0451  & 4.87 & 0.23 & 9.9 & 33.5 & 34.8 &   \cite{be00}\\
J1024$-$0719  & 5.16 & 0.35 & 9.6 & 33.7 & 34.6 &   \cite{su03}\\
J1744$-$1134  & 4.08 & 0.36 & 9.9 & 33.7 & 34.6 &   \cite{su03} \\
J1843$-$1113  & 1.85 & 1.97 & 9.5 & 34.8 & 34.2 &   \cite{ho04} \\
J1823$-$3021A  & 5.44 & 7.9 & 7.4 & 35.9 & 34.1 & \cite{bt97}  \\
J0024$-$7204F  & 2.62 & 4.8 & 8.8 & 35.1 & 33.8 & \cite{ro95}\\
J1730$-$2304  & 8.12 & 0.51 & 9.8 & 33.2 & 33.8 &   \cite{lo95} \\
J2322+2057  & 4.81 & 0.78 & 9.9 & 33.5 & 33.8 &      \cite{bt97} \\
J0711$-$6830  & 5.49 & 1.04 & 9.8 & 33.6 & 33.5 &  \cite{ba96} \\
J1910$-$5959D  & 9.04 & 4 & 8.2 & 34.7 & 33.5 & \cite{da02}  \\ 
J1944+0907  & 5.19 & 1.28 & 9.7 & 33.7 & 33.5 &   \cite{ch05} \\
J1721$-$2457  & 3.50 & 1.56 & 10.0 & 33.7 & 33.4 &   \cite{ed01} \\
J1905+0400  & 3.78 & 1.34 & 10.1 & 33.5 & 33.3 &  \cite{ho04}\\
J1801$-$1417  & 3.63 & 1.8 & 10.0 & 33.6 & 33.1 &  \cite{fa04} \\

\hline
\end{tabular}
\label{milli}
\end{table}
\vspace{2mm}
Optical GRP observations are currently limited to the Crab pulsar and likely to remain so for the foreseeable future.  The frequency of GRP events is very low with all other GRP emitters and only one of which (PSR J0540$-$6919) has been observed optically. If PSR J0540$-$6919 behaves in a similar manner to the Crab pulsar then the observed GRP rate of 0.001 Hz, about 1000 times lower than the Crab, will have a corresponding increase in observing time making optical observations unlikely in the near future. If PSR J0540$-$6919 shows a  3\% increase in optical flux  during GRP events then detecting this at the 3 $\sigma$ level would require over 100 hours of observation using the 8m VLT and an APD based detector. 

\vspace{2mm}

{\bf Millisecond Pulsars}
\vspace{1mm}

With the exception of UV observations of the  PSR J0437$-$4715 \cite{ka04} there are  no optical counterparts of any millisecond pulsar and in the case of PSR J0437$-$4715 the optical emission is probably thermal in origin. There have been a few upper limits and possible, but unconfirmed, counterparts. Until a number of optical millisecond pulsars has been observed, it will be difficult to even estimate the number of millisecond pulsars we can expect to observe optically . Table \ref{milli} shows those millisecond pulsars with the highest  $\dot{E}/d^2$, which which we regard as a starting point for selecting those pulsars suitable for optical study.

\vspace{2mm}

{\bf AXPs}
\vspace{1mm}

AXPs have typical periods in the range 8-12 seconds and consequently can be observed with conventional  CCDs or Low-Light-Level CCDs (L3CCD). They also are characterised by 
high magnetic fields indicating that polarisation will be an important component of any future instrument and programme of observations. However the number of known AXPs is  small and most have upper limits which are significantly lower than the detected flux from 4U 0142+61 making significant optical observations unlikely in the short term. In the near-IR recent observations of 4U 0142+61 \cite{wang06} are consistent with the existence of a cool (T $\approx$ 91 K) debris disk around the neutron star. We would expect, if this is confirmed, that the pulse fraction in the near-IR should be lower than in the optical wavebands. It is interesting to speculate whether similar features could be observed in young normal pulsars..

\vspace{2mm}

{\bf RRATs}
\vspace{1mm}

Observing RRATs in the optical will be a serious observational challenge requiring coordinated optical/radio-observations. For example PSRJ1819$-$1458 was observed for $\approx$ 1800 seconds with UltraCam which produced  upper magnitude limits of 15.1, 17.4 and 16.6 in $u'$, $g'$ and $i'$. If only those frames coincident with radio events were recorded then this would have produced 3$\sigma$ limits of 16.7, 18.8 and 18.1 respectively. If the link to AXPs is real then we must obtain  peak $i'$ magnitudes of about 24 or unrealistic exposure times of $\approx$ 71 hours using UltraCam. Even if we use a detector with zero read noise (e.g. L3CCDs or APDs) then our exposure times drop down to 57 hours, for a 3 $\sigma$ detection. Alternatively if, as with GRPs, the energy in the radio and optical transient are similar then we would expect observation times of roughly 10 hours with UltraCam to achieve a 3 $\sigma$ detection. Another way of looking at the RRAT detection probability is to scale according to the fractional on-time of each RRAT. PSR J1819$-$1458 is radiating for approximately 0.01\% of the time or 10 magnitudes down from a pulsar which is radiating continuously. In contrast PSR J1848$-$12 is a further factor of 100 weaker. 

\section{Conclusion}

Optical/infra-red observations of pulsars of all types are still in their relative infancy with very few objects detected. The likelihood is that in the next 5 years the number of pulsars with detected optical emission will have doubled and more significantly, new observations will include polarisation measurements.  With respect to high energy emission, it is only at optical energies where it is possible to measure, with reasonable accuracy, all electro-magnetic aspects of pulsar radiation. These later observations, combined with detailed numerical models should elucidate,  through geometric arguments, the structure of the emission zone for normal pulsars and hence provide stringent observational tests for the various models of pulsar emission. In the future ELTs with adaptive optics will enable pulsar observations down to 32-33 magnitude and importantly will enable spectra-polarimetry down to 29-30 magnitude.

Beyond this time into the ELT era there is the possibility of extending the number of optically observed pulsars (and pulsar types) dramatically. This will require detectors with :-
\begin{itemize} 
\item High-time resolution --- both frame based systems from 1 ms+ and pixel read-outs for $\tau < $ 1 ms
\item Medium size arrays --- wide-field operations are not required but nearby reference stars are needed - array sizes $>$ 32 $\times $ 32
\item Polarisation --- sensitive to polarisation changes - both circular and linear at the 1\% level
\item Energy Resolution --- Broad-medium band energy resolution - 1-10\%
\end{itemize}

\index{paragraph}

%
%

%
%

\printindex
\end{document}